# Branching of the vortex nucleation period in superconductor Nb microtubes due to inhomogeneous transport current

*R. O. Rezaev*[1,2], *E. A. Levchenko*[1], *V. M. Fomin*[3,*]

[1]National Research Tomsk Polytechnic University, Lenin av. 30, Tomsk, 634050, Russia
[2]National Research Nuclear University MePhI, Kashirskoye sh. 31, Moscow, 115409, Russia
[3]Institute for IntegrativeNanosciences, IFW Dresden, Helmholtzstraße 20, D-01069 Dresden, Germany
*Corresponding author; e-mail: v.fomin@ifw-dresden.de

## ABSTRACT

An inhomogeneous transport current, which is introduced through multiple electrodes in an open Nb microtube, is shown to lead to a controllable branching of the vortex nucleation period. The detailed mechanism of this branching is analyzed using the time-dependent Ginzburg-Landau equation. The relative change of the vortex nucleation period strongly depends on the geometry of multiple electrodes. The average number of vortices occurring in the tube in a nanosecond can be effectively reduced owing to the inhomogeneous transport current, what is important for noise and energy dissipation reduction in superconductor applications, e.g., for an extension of the operation regime of superconductor-based sensors to lower frequencies.

KEYWORDS: superconducting vortices, vortex dynamics, branching, vortex nucleation period, time-dependent Ginzburg-Landau equation, nanotechnology, computational physics

## I. INTRODUCTION

Control over the vortex dynamics in superconductors provides an efficient tool to study superconducting properties of materials and is an important issue for applications. In most cases a particular control technique is realized by the guidance of vortices through predefined trajectories [1-4], for example, by the rectification of the average vortex movement [5-8].The guidance usually is implemented through an array of artificial pinning sites [1-3, 9, 10], which lead to a redistribution of the local current density. Meanwhile, fabrication of curved superconductor structures provides another possibility to affect the current density distribution as a function of curvature. The most pronounced control over the vortex dynamics by this approach occurs when the curvature radius is of the order of the coherence length $\xi$ [11-13].

Recent technological advances have allowed for fabrication of rolled-up structures [14-16] that consist of InGaAs/GaAs/Nb layers. Nb shows mechanical homogeneity and its coherence length $\xi$ is comparable with thickness of the Nb layer in the rolled-up structures. Theoretical investigations have revealed a crucial role of the curved geometry of the rolled-up structures for the vortex dynamics [13]. A combined effect of pinning centers and a curved geometry has been demonstrated to strongly depend on the location of the pinning centers in a superconductor structure [17]. Because of correlated dynamics of vortices in different parts of a superconductor structure, the pinning centers have an impact on a long distance. This influence leads to the branching of a period [17] of vortex nucleation on different sides of an open superconducting Nb tube relative to the applied magnetic field, which is orthogonal to the axis of the tube. The magnitude of the branching is determined by positions and strength of pinning centers and characterizes a certain pattern of the vortex dynamics.

We suggest to manipulate the branching in such a superconductor structure, where the current density distribution can be controlled locally near the positions of vortex nucleation. A possible implementation of this idea is to fabricate multiple mutually isolated electrodes, through which the transport current is introduced into the structure. A gap between any two neighboring electrodes leads to a transport current discontinuity. In a planar structure in the presence of a homogeneous magnetic field, such discontinuities become points of vortex nucleation or denucleation. However, in a curved structure, such a geometry can be chosen, that in the vicinity of the joints between the electrodes, the component of the magnetic field normal to the surface of the structure vanishes. In this case, discontinuities of electrodes do not act as points of vortex nucleation.

In the present work, we evaluate the effect of a two-component electrode on the vortex dynamics in open Nb tubes and show a possibility to efficiently control their characteristics considering branching of the vortex nucleation period as an example.

## II. THEORETICAL MODEL

We consider a Nb superconductor open tube [13] of radius $R$ and length $L$ (Fig. 1a) with three paraxial contacts on the edges of the slit. The "red" contact plays the role of an input contact, through which the transport current $I_{in}$ enters. Through the "blue" contact, the current $I_{out}$ leaves the tube. Through the "green" contact, the additional transport current $I_{control}$ enters; it might be positive (in) or negative (out). Its role is to dynamically control the vortex nucleation. At each instant, the sum of all three currents satisfies the condition of continuity:

$$I_{in} + I_{control} + I_{out} = 0. \tag{1}$$

The system is placed in a magnetic field $\mathbf{B} = -B\mathbf{e}_z$ (Fig. 1a), which induces Meissner currents circulating at each half-tube [13]. The total current, which is a sum of Meissner and transport currents, is shown schematically in Fig. 1a by the black lines on the "front" half of the tube. Two of three currents in Eq. (1) are independent, which allows for different regimes of control over vortex dynamics. In the present work, we keep $I_{in}$ constant and change $I_{control}$.

Our model is based on the time-dependent Ginzburg-Landau (GL) equation [18] for the order parameter $\psi$ in the dimensionless form [2]

$$\frac{\partial \psi}{\partial t} = (\nabla - i\mathbf{A})^2 \psi + 2\kappa^2(1 - |\psi|^2)\psi, \tag{2}$$

$$(\nabla - i\mathbf{A})\psi|_{n,boundary} = 0, \qquad \left(\nabla - i\left[\mathbf{A} + \frac{\mathbf{j}}{|\psi|^2}\right]\right)\psi\bigg|_{n,electrodes} = 0, \tag{3}$$

where dimensionless variables are determined in Refs. [2, 13, 17], $\kappa$ is the GL parameter, $\nabla$ is the gradient operator. The temperature dependent values of magnetic field considered in the present work belong to the 10 mT-range for the temperature $T=0.95T_c$, where $T_c$ is the critical temperature. The empirical law, which represents the thermodynamic critical magnetic field as a function of temperature (see Eq. (1.2) of Ref. [18]):

$$B_c(T) \approx B_c(0)[1 - (T/T_c)^2], \tag{4}$$

in view of Eqs. (5.18) and (4.62) of Ref. [18] is applicable for the first ($B_{c1}$) and second ($B_{c2}$) critical magnetic fields and thereby provides an estimate of the magnetic field range, where vortices occur at different temperatures.

The normal (with respect to the cylindrical surface) component of the magnetic field, according to Fig. 1a, is $B_n = B\sin(\varphi)$. The vector potential is taken in the Landau gauge: $\mathbf{A} = A\mathbf{e}_x$, $A = -By$. Transport currents in Eq. (1) are obtained from the current densities $\mathbf{j}$ in the boundary conditions of Eq. (3) by integrating along the appropriate contacts. The thickness of the superconductor wall, which is necessary to calculate the electron mean free path, constitutes 50 nm [13]. The typical tube radius and length under consideration are more than 10 times larger than the thickness. With these parameters, it is safe to neglect the effect of an induced magnetic field and to use a 2D approximation. Our numerical simulations show that the induced magnetic field is as low as 1% of the applied magnetic field. In our numerical simulation at $0.95T_c$, the coherence length constitutes $\xi = 56$ nm [13].

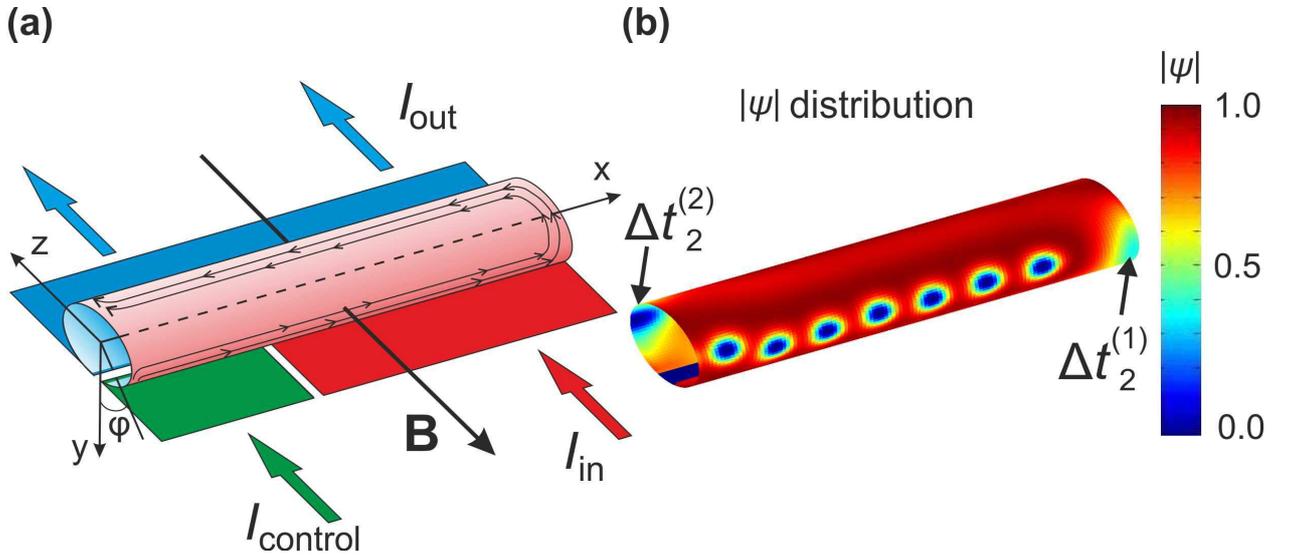

**Figure 1**. (**a**) Scheme of an open tube. Three mutually isolated electrodes are displayed by the red, blue and green areas. The *x*-axis is the axis of the cylindrical coordinate frame $(\rho, \varphi, x)$. The angle $\varphi$ is counted from the positive direction of the *y*-axis, which passes through the middle of the slit. $\mathbf{B} = -B\mathbf{e}_z$ is the applied magnetic field. (**b**) An example of the distribution of the amplitude of the order parameter $|\psi|$ at the magnetic field $B = 10$mT for a tube of radius $R = 500$ nm and length $L = 3.5\mu$m (After Ref. 13). Vortices nucleate at the front and back halves of the tube with periods $\Delta t_2^{(1)}$ and $\Delta t_2^{(2)}$, correspondingly, move along the *x*-axis and denucleate at the edge opposite to the point of nucleation.

## III. BRANCHING OF THE VORTEX NUCLEATION PERIOD

Two full-side electrodes generate symmetrical vortex dynamics – both halves of the tube demonstrate the same characteristic times [13]. In the present work, the period of vortex nucleation $\Delta t_2$ is considered to characterize the vortex dynamics, since it has a more pronounced dependence on the current density, than the duration of the vortex motion through the tube $\Delta t_1$. In what follows, we use the denotation $\Delta t_2^{(1)}$ for the half-tube with input and control electrodes (front side in Fig. 1a) and $\Delta t_2^{(2)}$ for the half-tube containing the output electrode (back side in Fig. 1a). For a tube of radius $R = 400$ nm with two full-side electrodes carrying the current $I = $ 1.7 mA at $B = 10$ mT, the nucleation period $\Delta t_2^{(1)} = \Delta t_2^{(2)} = \Delta t_2^{(sym)} = 1.2$ ns (this value is shown in Fig. 2 by the black solid line). However, introducing the control electrode of length $L_{control} = 20\xi$ into the tube of length $L = 60\xi$ (so that the input electrode is of length $L_{in} = 40\xi$) violates the inversion symmetry of the modulus of order parameter with respect to the geometric center of the tube. A similar mechanism occurs in vortex ratchets with a pinning potential, which lacks centrosymmetry [19].

In Fig. 2, the dependence of $\Delta t_2$ on the control current demonstrates a different behavior for each half-tube. We perform simulations for two radii of the tube. In Fig. 2, the blue line corresponds to $R = 600$ nm and the red one to $R = 400$ nm. An increase of the radius shifts both $\Delta t_2^{(1)}$ and $\Delta t_2^{(2)}$ curves upwards. As seen from Fig. 2, a difference between vortex nucleation periods $\Delta t_2^{(1)}$ and $\Delta t_2^{(2)}$ by a factor of as large as 3 is achieved. This difference, as follows from our simulation, is determined by the control current. Because of inhomogeneity of the current density distribution over the whole tube, a variation of the control current at one side of the tube leads to a change of the vortex nucleation period at the opposite side. A further decrease of the control current (beyond the values given in Fig. 2) leads to an infinite rise of $\Delta t_2^{(2)}$, what means that vortex nucleation is blocked on one side of the tube and the dynamics occurs completely on another side of the tube.

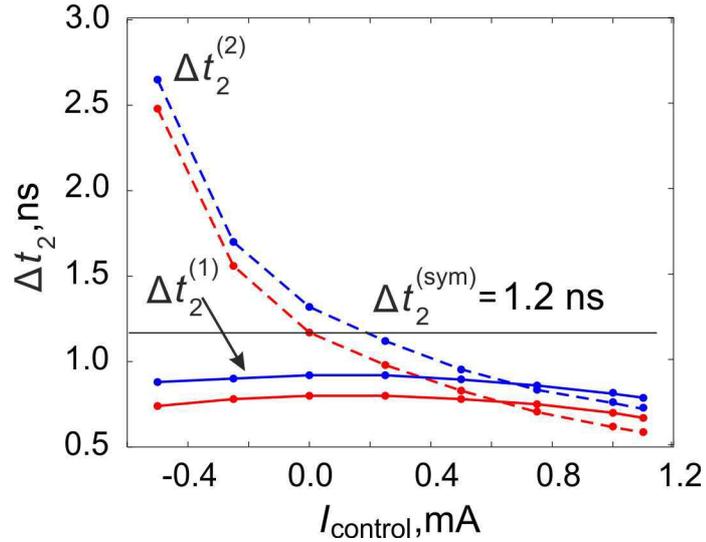

**Figure 2**. Vortex nucleation period as a function of the control current. The blue line corresponds to the tube of radius $R=600$ nm and the red line to $R = 400$ nm, the length of both tubes $L = 60\xi = 3360$ nm, the length of the control electrode $L_{control} = 20\xi = 1120$ nm. The input current is equal to 1.7 mA. $\Delta t_2^{(2)}, \Delta t_2^{(1)}$ are periods of vortex nucleation at opposite sides of a tube (see a detailed definition in the text). A black solid line represents the case of full-side electrodes, when input and output currents are equal by modulus to 1.7 mA for $R=400$ nm. In all cases the magnetic field $B = 10$ mT.

A variation of the control current from negative to positive values shows a crossover between $\Delta t_2^{(2)}$ and $\Delta t_2^{(1)}$ functions. For the radius $R = 600$ nm the crossover occurs at $I_{control} = 0.70$ mA, and for $R = 400$ nm it occurs at $I_{control} = 0.62$ mA. The dependence of $\Delta t_2^{(2)}$ on $I_{control}$ after passing the crossover point is less expressed than before: $\Delta t_2^{(2)}$ is even reduced as compared to $\Delta t_2^{(1)}$.

From the practical point of view, it is interesting to evaluate the average number of vortices in a nanosecond when the dynamics occurs only at one side of the tube. In the tube with radius $R = 400$ nm, the length of the control electrode $L_{control} = 1120$ nm and $I_{control} = -0.5$ mA, which is analyzed here as a typical case, the characteristic times $\Delta t_2^{(2)} \to \infty$ and $\Delta t_2^{(1)} \approx 0.7$ ns result in the average number of vortices $n_v \approx 1.43$ in a nanosecond. In a tube with full-side electrodes, $n_v \approx 1.67$ in a nanosecond, so that the relative difference in the average number of vortices constitutes ~ 15% for the same input current. Thus, using inhomogeneous transport current in the tube leads to an effective decrease of the average number of vortices. Such a decrease plays a crucial role in noise and energy dissipation reduction for numerous superconductor applications [5, 19]. In fact, the property $\Delta t_2^{(2)} \to \infty$ implies that using multiple electrodes allows for vortex removal from certain regions of a superconductor sample, which is of practical interest, for example, in order to suppress the 1/$f$-noise due to the activated hopping of trapped vortices and thus to extend the operation regime of superconductor-based sensors to lower frequencies [5].

## IV. CURRENT DENSITY MODIFICATION BY THE CONTROL ELECTRODE

Vortex dynamics occurs under Magnus force [**j, B**] and linearly depends on the local current density [18]. Geometry of boundaries in mesoscopic superconductors determines the points of nucleation, denucleation and possible location of vortices [20], and the corresponding current density distribution leads to the non-linear character of vortex motion as a function of the control current. The branching of the vortex nucleation period shown in Fig. 2 is an example of non-linear dynamics. An interpretation of such a behavior of vortices requires to analyze the current density. In Fig. 3, the key mechanism of the difference between vortex dynamics in the cases with (Fig. 3b) and without (Fig. 3a) a control electrode is illustrated. Near the points of vortex nucleation for both halves of the tube, the current density components are listed in the rectangles. The black and light grey thin arrows on each panel in Fig. 3 point to the same positions on the tube. As clearly seen, the main reason for the difference in vortex dynamics is a change of the current density component $j_s$ along the azimuthal direction (a corresponding coordinate $s = R\varphi$ is defined through the azimuthal angle $\varphi$ shown in Fig. 2). From comparison of two bottom panels in Fig. 3, a decrease of this component at the points of vortex nucleation specified by the black thin arrows on the side of the output electrode ($j_s^a - j_s^b = 2.49 - 2.39 = 0.1$) is higher than the current density increase ($j_s^b - j_s^a = 2.42 - 2.39 = 0.03$) at the points shown by light grey thin arrows on the side of the input electrode. At the same time, the $j_x$ components are similar to each other in both cases.

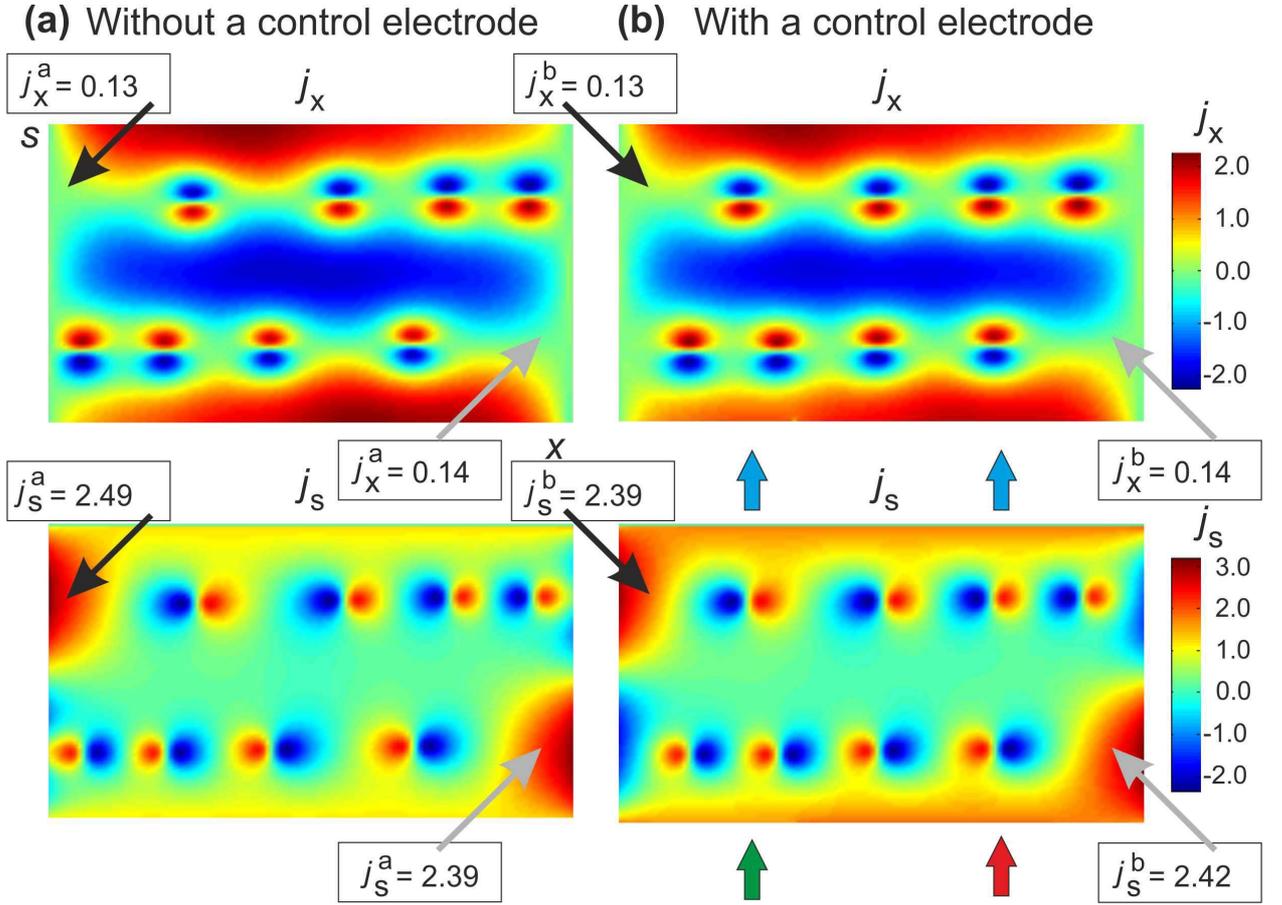

**Figure 3**. Distributions of the $j_x$ and $j_s$ components of the current density over the surface of a tube in the $(x, s)$-coordinates, where $x$ is defined in Fig. 1a and $s = R\varphi$. Radius of the tube is $R = 400$ nm. The magnetic field is $B = 8$mT. (**a**) Two full-side electrodes are attached: input and output ones with the transport current $I = 1.7$ mA. (**b**) Input, control and output electrodes of length $L_{in} = 40\xi$, $L_{control} = 20\xi$ and $L_{output} = 60\xi$, correspondingly, are attached. The currents are $I_{in} = 1.7$ mA and $I_{control} = 0.75$ mA. The green, red and blue arrows show the control, input and output currents.

## V. EFFECT OF THE MAGNETIC FIELD AND THE LENGTH OF THE CONTROL ELECTRODE

The vortex nucleation period is a decreasing function of the magnetic field as shown in Fig. 4a for two cases, without and with the applied control current 0.5 mA. For higher values of the control current, the difference between $\Delta t_2^{(2)}$ and $\Delta t_2^{(1)}$ is less pronounced, than for lower values of the control current (see, for instance, a crossover point in Fig. 2 for a fixed magnetic field).

For a lower magnetic field, the absolute value of the difference between the two periods is larger, than for a higher magnetic field. For example, $\left|\Delta t_2^{(2)} - \Delta t_2^{(1)}\right| \approx 0.6$ ns for $B = 8$mT and $\left|\Delta t_2^{(2)} - \Delta t_2^{(1)}\right| \approx 0.3$ ns for $B = 20$ mT for the case without a control current. However, the relative difference $\delta_{21} = \left|\Delta t_2^{(2)} - \Delta t_2^{(1)}\right|/\Delta t_2^{(2)} \approx 0.4$ is varying at most by 5% within the considered range of magnetic fields. Since the $\Delta t_2^{(2),(1)}(B)$ curves saturate with the magnetic field growth [13], the relative difference $\delta_{21}$ keeps almost steady for the whole range of magnetic fields, where vortex dynamics occur. Qualitatively, the functions $\Delta t_2^{(2),(1)}(B)$ have the same shape for the control current $I_{control}$ = 0.5 mA (see the blue lines in Fig. 4a), for which the relative difference $\delta_{21}$ is about 0.1.

The relative difference $\delta_{21}$ strongly depends on the length of the control electrode, as shown in Fig. 4b. For $I_{conrtol} = 0$ mA and the control electrode of length $L_{control} = 10\xi$ the relative difference is $\delta_{21} \approx 0.14$, while for the length $L_{control} = 30\xi$ the relative difference is significantly larger: $\delta_{21} \approx 0.52$. The main reason for this dramatic change is a decrease of $\Delta t_2^{(1)}$ more than twice, while $\Delta t_2^{(2)}$ only slightly depends on $L_{control}$. The decrease of $\Delta t_2^{(1)}$ in Fig. 4b results from the change of the length of the input electrode, correlated with the length of the control electrode. In particular, the input current density $j_{in}$ rises in the vicinity of the vortex nucleation point, what leads to an effective reduction of the potential barrier (see Fig. 6 of Ref. [13]). The vortex nucleation period $\Delta t_2^{(1)}$ decreases, while the corresponding period $\Delta t_2^{(2)}$ at the opposite side of the tube remains practically unchanged, since the output current density is not affected by changing the length of the input electrode.

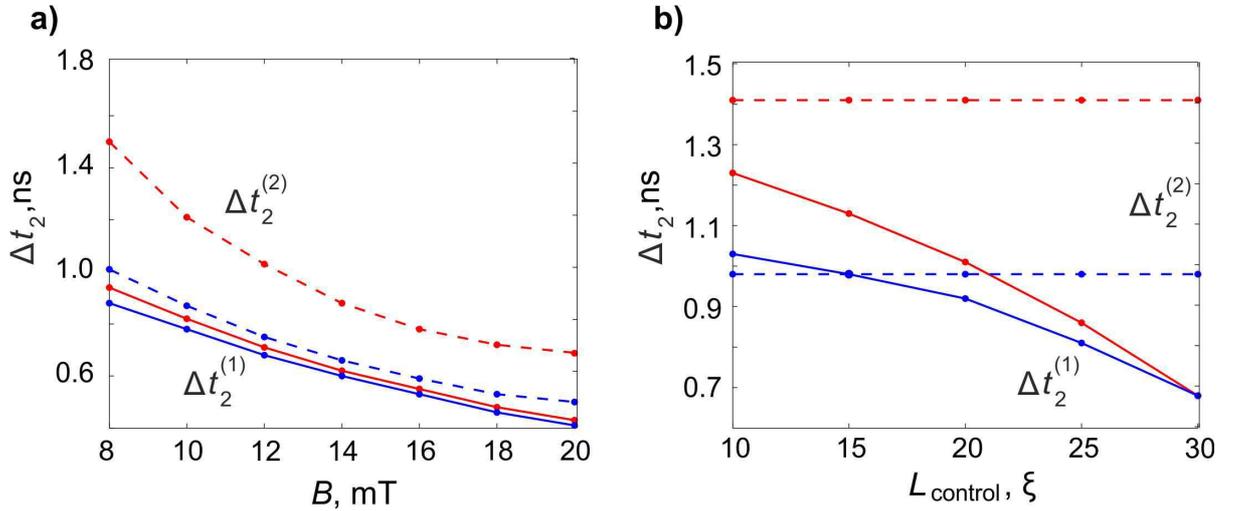

**Figure 4**. $\Delta t_2^{(2)}$ ($\Delta t_2^{(1)}$) are denoted by dashed (solid lines), correspondingly. (**a**) The vortex nucleation period as a function of the magnetic field for the tube of radius $R = 400$ nm, the input current $I_{in} = 1.7$ mA and the control electrode of length $L_{control} = 20\xi$. (**b**) The vortex nucleation period as a function of the control electrode length for the tube of radius $R = 400$ nm and the input current $I_{in} = 1.5$ mA at the magnetic field $B = 10$ mT. In both panels, the red lines correspond to the case without a control current and the blue lines are drawn for $I_{control} = 0.5$ mA.

## VI. SUMMARY

In conclusion, the interplay of a curved geometry with an inhomogeneous transport current determines a specific current density distribution, which destroys the inversion symmetry of the order parameter with respect to the geometric center of the tube and leads to a branching of the vortex nucleation period. In particular, using the appropriate electrode arrangement, the vortex dynamics can be blocked on one side of the tube. The relative change of the vortex nucleation period weakly depends on the magnetic field in the range, where vortex dynamics occur for the considered control currents. However, it strongly depends on the length of the control electrode. The proposed method allows for tuning the frequency of vortex generation on different parts of the tube and provides a reduction of the average number of vortices in a nanosecond what is important for noise and energy dissipation reduction in superconductor applications, for instance, for an extension of the operation regime of superconductor-based sensors to lower frequencies.

## ACKNOWLEDGEMENT

Discussions with Oliver G. Schmidt, Hermann Suderow, Sören Lösch and Danilo Bürger are gratefully acknowledged. R. O. R. acknowledges kind hospitality during his stay at the IIN, IFW-Dresden. The work was supported by the bilateral BMBF-Russia research grant 01DJ13009, by the COST Action MP1201 'Nanoscale Superconductivity' under the grant COST-STSM-MP1201-24449, by the Program 'Science' under contract No. 1.676.2014/ K (Russia) and by the collaborative project with ZIH of Dresden University of Technology.